\documentclass[twocolumn,showpacs,preprintnumbers,amssymb]{revtex4}
\usepackage{graphicx, epsfig, amssymb} 
\usepackage{amsmath, amsfonts}
\usepackage{bm} 

\def\be{\begin{equation}}
\def\ee{\end{equation}}
\def\beq{\begin{eqnarray}}
\def\eeq{\end{eqnarray}}

\begin{document}

\centerline{}
\title{On the instability of Reissner-Nordstr\"om black holes in de Sitter backgrounds}

\author{Vitor Cardoso}
\email{vitor.cardoso@ist.utl.pt}
\affiliation{CENTRA, Departamento de F\'{\i}sica, Instituto Superior T\'ecnico,
Av. Rovisco Pais 1, 1049-001 Lisboa, Portugal \& \\
Department of Physics and Astronomy, The University of Mississippi,
University, MS 38677-1848, USA
}
\author{Madalena Lemos}
\email{madalena.dal@gmail.com}
\affiliation{CENTRA, Departamento de F\'{\i}sica, Instituto Superior T\'ecnico,
Av. Rovisco Pais 1, 1049-001 Lisboa, Portugal
}
\author{Miguel Marques}
\email{miguel.e.marques@gmail.com}
\affiliation{CENTRA, Departamento de F\'{\i}sica, Instituto Superior T\'ecnico,
Av. Rovisco Pais 1, 1049-001 Lisboa, Portugal
}
\date{\today}

\begin{abstract}
Recent numerical investigations have uncovered a surprising result: Reissner-Nordstr\"om-de Sitter black holes are unstable for spacetime dimensions larger than 6. Here we prove the existence of such instability analytically, and we compute the timescale in the near-extremal limit. We find very good agreement with the previous numerical results. Our results may me helpful in shedding some light on the nature of the instability.
\end{abstract}


\maketitle
\newpage
\section{Introduction}
In physics, stability of a given configuration (solution of some set of equations), 
is a useful criterium for relevance of that solution. Unstable configurations are likely not to be realizable in practice, 
and represent an intermediate stage in the evolution of the system. Nevertheless, the instability itself is of great interest, since an understanding 
of the mechanism behind it may help one to better grasp the physics involved. In particular, it is of interest to be able to predict which other systems display similar instabilities, or even have a deeper understanding of the physics behind the instability
(why is the system unstable? is there some fundamental principle behind the instability?).

In General Relativity, the Kerr family exhausts the black hole solutions to the electro-vac Einstein equations.
Kerr black holes are stable, and can therefore describe astrophysical objects.
However, there are many instances of instabilities afflicting objects with an event horizon, such as the Gregory-Laflamme \cite{Gregory:1993vy}, the ultra-spinning \cite{Emparan:2003sy} or superradiant instabilities \cite{Cardoso:2004nk} 
and other instabilities of higher-dimensional black holes in alternative theories
\cite{Dotti:2005sq,Takahashi:2009dz}(for a review see Ref. \cite{Harmark:2007md}).

Konoplya and Zhidenko (hereafter KZ) recently studied small perturbations in the vicinity
of a charged black hole in de Sitter background, a Reissner-Nordstr\"om de Sitter black hole (RNdS) \cite{Konoplya:2008au}. 
Their (numerical) results show that when the spacetime dimensionality $D>6$, the spacetime is unstable, provided the charge is larger than a given threshold, determined by KZ for each $D$. Because the results are so surprising (the mechanism behind it is not yet understood), we set out to
to investigate this instability and hopefully understand it better. Our results can be summarized as follows: (i) we can prove analytically the existence of unstable modes for charge $Q$ higher than a certain threshold. (ii) in the near-extremal regime, we are able to find an explicit solution for the unstable modes, determining the instability timescale analytically.
We hope that our incursion in this topic helps to better understand the physics at work.
\section{Equations}
This work focuses on the higher dimensional RNdS geometry,
described by the line element
\be
ds^2 = -f\, dt^2 + f^{-1}\, dr^2 + r^2 d\Omega_n^2\,, 
\label{lineelement}
\ee
where $d\Omega_n^2$ is the line element of the $n$ sphere and
\be
f=1-\lambda\,r^2-\frac{2M}{r^{n-1}}+\frac{Q^2}{r^{2n-2}}\,.
\ee
the background electric field is $E_0=q/r^n$, with $q$ the electric charge. The quantities $M$ and $Q$ are related to the physical mass {\cal M} and charge $q$ of the black hole \cite{Kodama:2003kk}, and $\lambda$ to the cosmological constant.
The spacetime dimensionality is $D=n+2$.

The above geometry possesses three horizons: the black-hole Cauchy horizon at $r=r_a$, the black hole event horizon is at $r=r_b$ and the
cosmological horizon is at $r = r_c$, where $r_c > r_b>r_a$, the only real, positive zeroes of $f$.
For convenience, we set $r_b=1$, i.e., we measure all quantities in terms of the event horizon $r_b$.
We thus get
\be
2M=1+Q^2-\lambda\,,
\ee
Furthermore, we can also write
\be
\lambda=\frac{r_c^{-4-n}(r_c^{n+2}-r_c^3)(r_c^{n+2}-Q^2r_c^3)}{r_c^{n+2}-r_c}\,.
\ee
For a fixed $r_c$ and spacetime dimension $D$, the existence of a regular event horizon imposes that the charge 
$Q$ must be smaller than a certain value $Q_{\rm ext}$. With our units this maximum charge is
\be
Q_{\rm ext}^2=\frac{r_c^n\left(-2r_c+(n+1)r_c^n-(n-1)r_c^{n+2}\right)}{-r_c\left(r_c(n+1)-2nr_c^n+(n-1)r_c^{2n+1}\right)}\,.\label{Qext}
\ee
Gravitational perturbations of this spacetime couple to the electromagnetic field, and were completely characterized by Kodama and Ishibashi \cite{Kodama:2003kk}.
They can be reduced to a set of two second order ordinary differential equations of the form,
\be
\frac{d^2}{dr_*^2}\Phi_\pm+\left(\omega^2-V_{S\pm}\right)\Phi_\pm=0\,,\label{we}
\ee
where the tortoise coordinate $r_*$ and the potentials $V_{S\pm}$ are defined through 
\be
r_* \equiv \int f^{-1}\, dr\,,\quad V_{S\pm} =\frac{fU_\pm}{64r^2H_\pm^2}\,.
\label{tortoise}
\ee
We have 
\beq
H_+&=&1-\frac{n(n+1)}{2}\delta x\,,\\
H_-&=&m+\frac{n(n+1)}{2}(1+m\delta)x\,,
\eeq
and the quantities $U_\pm$ are given by
\begin{widetext}
\beq
U_+ &=& \left[-4 n^3 (n+2) (n+1)^2 \delta^2 x^2-48 n^2 (n+1) (n-2) \delta x
\right.\nonumber\\
&&\left.   -16 (n-2) (n-4)\right] y
  -\delta^3 n^3 (3 n-2) (n+1)^4 (1+m \delta) x^4
\nonumber\\
&&   +4 \delta^2 n^2 (n+1)^2 
   \left\{(n+1)(3n-2) m \delta+4 n^2+n-2\right\} x^3
\nonumber\\
&&   +4 \delta (n+1)\left\{
   (n-2) (n-4) (n+1) (m+n^2 K) \delta-7 n^3+7 n^2-14 n+8
   \right\}x^2
\nonumber\\
&&  + \left\{16 (n+1) \left(-4 m+3 n^2(n-2) K\right) \delta
     -16 (3 n-2) (n-2) \right\}x
\nonumber\\
&&    +64 m+16 n(n+2) K,\\
U_- &= & \left[-4 n^3 (n+2) (n+1)^2 (1+m \delta)^2 x^2
      +48 n^2 (n+1) (n-2) m (1+m \delta) x  \right.
\nonumber\\
&& \left.  -16 (n-2) (n-4) m^2\right] y
     -n^3 (3 n-2) (n+1)^4 \delta (1+m \delta)^3 x^4
\nonumber\\
&& -4 n^2 (n+1)^2 (1+m \delta)^2 
     \left\{(n+1)(3 n-2) m \delta-n^2\right\} x^3
\nonumber\\
&&  +4 (n+1) (1+m \delta)\left\{ m (n-2) (n-4) (n+1) (m+n^2 K) \delta
  \right. \nonumber\\
&& \left. \quad  +4 n (2 n^2-3 n+4) m+n^2 (n-2) (n-4) (n+1)K \right\}x^2
\nonumber\\
&&  -16m \left\{ (n+1) m \left(-4 m+3 n^2(n-2) K\right) \delta
\right.\nonumber\\
&&\left.  +3 n (n-4) m+3 n^2 (n+1) (n-2)K \right\}x
\nonumber\\
&&      +64 m^3+16 n(n+2)m^2 K.
\eeq
\end{widetext}
The variables $x,y$ and parameters $\mu,m$ are defined through
\beq
x&\equiv& \frac{2M}{r^{n-1}}\,,\quad y\equiv \lambda\,r^2\,,\\
\mu^2&\equiv& M^2+\frac{4m Q^2}{(n+1)^2}\,,\quad m\equiv k^2-n K\,,
\eeq
and the quantity $\delta$ is implicitly given by $\mu =(1+2 m \delta)M$.
Note that the following relations holds $Q^2=(n+1)^2M^2 \delta( 1+m\delta)$.

Note also that for the spacetime considered in this paper $K=1$, which means that the eigenvalues $k^2$ are given by $k^2=l(l+n-1)$, where $l$ is the angular quantum number, that gives the multipolarity of the field.
\begin{figure}[htpb!]
\includegraphics[width=9cm]{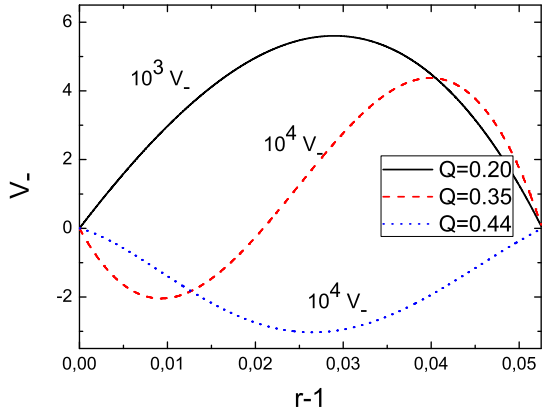}
\caption{\label{fig:pots} Behavior of $V_{-}$ for different parameters, for $D=8$. Here we fix the event horizon at $r_b=1$, and the cosmological horizon at
$r_c=1/0.95$. We consider $l=2$ modes and three different charges, $Q=0.2,0.35,0.44$.} 
\end{figure}
The behavior of the potentials varies considerably over the range of parameters. In Fig.~\ref{fig:pots} we show
$V_{-}$ for $D=8$, $r_c=1/0.95$, $l=2$ and three different values of the charge, $Q=0.2,0.35,0.44$.
\section{\label{sec:negativeV}A criterium for instability}
A sufficient (but not necessary) condition for the existence of an unstable mode
has been proven by Buell and Shadwick \cite{buell} and is the following,
\be
\int_{r_b}^{r_c}\frac{V}{f}dr<0\,.\label{Vcriterium}
\ee
\begin{figure}[htpb!]
\includegraphics[width=9cm]{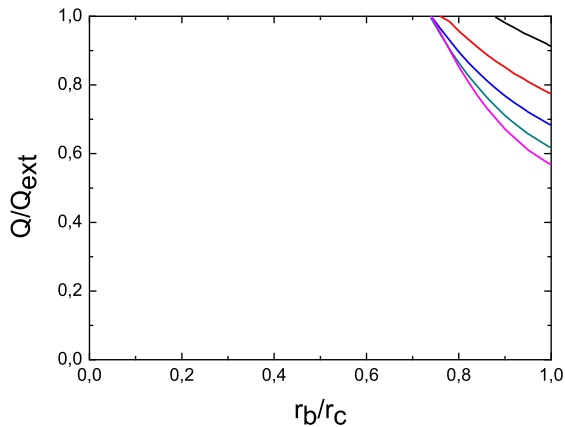}
\caption{\label{fig:threshold} The parametric region of instability in $Q/Q_{\rm ext}-r_b/r_c$ coordinates,
 according to criterim (\ref{Vcriterium}), for $l=2$. Top to bottom, $D=7,8,9,10,11$.} 
\end{figure}
The instability region is depicted in figure \ref{fig:threshold} for several spacetime-dimension $D$, which can be compared with the numerical results by KZ, their figure 4. It is apparent that condition (\ref{Vcriterium}) very accurately describes the numerical results for $r_b/r_c \sim 1$, 
a regime we explore below in Section \ref{sec:ne}. As one moves away from extremality criterium (\ref{Vcriterium}) is just too restrictive.
An improved analysis and refined criterium would be necessary to describe the whole range of the numerical results.
Nevertheless, figure \ref{fig:threshold} is very clear: higher-dimensional ($D>6$) RNdS black holes are unstable for a wide range of parameters.
\section{\label{sec:ne}An exact solution in the Near Extremal RNdS Black Hole}

Let us now specialize to the near extremal RNdS black hole, which we define as the spacetime for which the cosmological horizon $r_c$
is very close (in the $r$ coordinate) to the black hole horizon $r_b$, i.e.
$\frac{r_c-r_b}{r_b}\ll1$. The wave equation in this spacetime can be solved exactly, in terms of hypergeometric functions \cite{Cardoso:2003sw}.
The key point is that the physical region of interest (where the boundary conditions are imposed), lies between $r_b$ and $r_c$. Thus,
\be
f \sim 2\kappa_b\frac{(r-r_b)(r_c-r)}{r_c-r_b}\,,
\label{approximation2}
\ee
where we have introduced the surface gravity $\kappa_b$ associated with the
event horizon at $r = r_b$, as defined by the relation $\kappa_b = \frac{1}{2} 
df/dr_{r=r_b}$. For near-extremal black holes, it is approximately
\be
\kappa_b \sim \frac{(r_c-r_b)(n-1)}{2r_b^2}\left(1-nQ^2\right)\,.
\ee
In this limit, one can invert the relation $r_*(r)$ 
of (\ref{tortoise}) to get
\be
r= \frac{r_c e^{2\kappa_b r_*}+r_b}{1+e^{2\kappa_b r_*}}\,.
\label{rtortoise}
\ee
Substituting this on the expression (\ref{approximation2})
for $f$ we find 
\be
f = \frac{(r_c-r_b)\kappa_b}{2\cosh{(\kappa_b r_*)}^2}\,.
\label{approximation3}
\ee
As such, and taking into account the functional form of the potentials
for wave propagation, we see that for
the near extremal RNdS black hole the wave equation (\ref{we}) is of the form
\be
\frac{d^{2} \Phi(\omega,r)}{d r_*^{2}} +
\left\lbrack\omega^2-\frac{V_0}{\cosh{(\kappa_b r_*)}^2}\right\rbrack
\Phi(\omega,r)=0 \,,
\label{waveequation2}
\ee
with
\be
V_0=\frac{(r_c-r_b)\kappa_b}{2}\frac{V_{S\pm}(r_b)}{f}
\ee
The potential in (\ref{waveequation2}) is the well known 
P\"oshl-Teller potential \cite{teller}. The solutions to 
(\ref{waveequation2})
were studied and they are of the hypergeometric type, 
(for details see Refs. \cite{Ferrari:1984zz,Berti:2009kk}).
It should be solved under appropriate boundary conditions:
\beq
\Phi & \sim & e^{-i\omega r_*} \,,\quad r_* \rightarrow -\infty \\
\Phi &\sim & e^{i\omega r_*}  \,,\hskip 6mm r_* \rightarrow \infty. 
\label{behavior1}
\eeq
These boundary conditions impose a non-trivial condition on
$\omega$ \cite{Ferrari:1984zz,Berti:2009kk}, and those that satisfy both simultaneously
are called quasinormal frequencies. For the P\"oshl-Teller potential
one can show \cite{Ferrari:1984zz,Berti:2009kk} that they are given by
\be
\omega=\kappa_{b} \left [ -\left(j+\frac{1}{2}\right)i+
\sqrt{\frac{V_0}{\kappa_b^2}-\frac{1}{4}} \right ]\,, \quad \,j=0,1,...\,.
\label{solution}
\ee
\begin{table}[h]
\caption{\label{tab:exponent} The threshold of instability for near-extremal RNdS black holes (i.e., black holes for which the cosmological and event horizon almost coincide) for $l=2$ modes.
We show the prediction from the exact, analytic expression obtained in the near extremal limit (\ref{analytic}), which we label $Q/Q_{\rm ext}^{N}$  
and the one from criterium (\ref{Vcriterium}) which we label as $Q/Q_{\rm ext}^{V}$. Both these results are compared to the numerical results by KZ.}
\begin{tabular}{ccccccc}
\hline 
&\multicolumn{5}{c}{$D$}\\ \hline
                      &7    &8     &9    &10   &11&$D\to \infty$\\ 
$Q/Q_{\rm ext}^{N}$   &0.913&0.774 &0.683&0.617&0.567&$\sqrt{2/D}$\\
$Q/Q_{\rm ext}^{V}$   &0.913&0.775 &0.684&0.618&0.568&$\sqrt{2/D}$\\ 
$Q/Q_{\rm ext}^{Num}$ &0.94&0.78 &0.68&0.61&0.55&---\\ \hline
\end{tabular}
\end{table}
We conclude therefore that an instability is present whenever $V_0$ is negative.
The threshold of stability in the near-extremal regime is therefore given by
\be
\frac{V_{S\pm}(r_b)}{f}=0\,,\label{analytic}
\ee
%
The expression for $V_{S\pm}(r_b)/f$ is lengthy, and we won't present it here. The values of the charge $Q/Q_{\rm ext}$
that satisfy the condition above are given in Table I (for $l=2$), and compared to the prediction from the analysis in Section \ref{sec:negativeV}, criterium (\ref{Vcriterium}). The agreement is excellent. Furthermore,
we compare these predictions against the numerical results by KZ, extrapolated to the extremal limit ($\rho=1$ in KZ notation). The agreement is remarkable.
\section{Conclusions}
We have shown analytically that charged black holes in de Sitter backgrounds are unstable for a wide 
range of charge and mass of the black hole, confirming previous numerical studies \cite{Konoplya:2008au}.
The stability properties of the extremal $D=6$ black hole remain unknown. Our methods and results and inconclusive at this precise point, further dedicated investigations would be necessary.

Our analytical result in the near-extremal regime could be used to investigate further the nature of this instability, something we have not attempted to do here. A possible refinement concerns the large-$D$ limit of the instability, where 
it {\it could} be possible to find an analytical expression throughout all range of parameters. We have in mind results and techniques similar to those of Kol and Sorkin \cite{Kol:2004pn}. It would also be interesting to investigate the stability properties, using this or other techniques, of near-extremal Kerr-dS black holes, which have recently been conjectured to have an holographic description \cite{Anninos:2009yc}.
\section*{Acknowledgements}
We warmly thank Roman Konoplya and Alexander Zhidenko for useful correspondence and for sharing their numerical results with us. 
This work was partially funded by Funda\c c\~ao para a Ci\^encia e Tecnologia (FCT)-Portugal through projects
PTDC/FIS/64175/2006, PTDC/ FIS/098025/2008, PTDC/FIS/098032/2008 and CERN/FP/109290/2009.

\end{document}